\documentclass[prb,aps,reprint,amsmath,amssymb,superscriptaddress,longbibliography]{revtex4-2}

\usepackage{graphicx}
\usepackage{dcolumn}
\usepackage{bm}
\usepackage{xcolor}

\begin{document}

\title{Accidental accuracy and vertex corrections in $GW$: Exact benchmarks for the extended Hubbard model}

\author{Michael O. Atambo}
\email{michael.atambo@tukenya.ac.ke}
\affiliation{Department of Physics, Earth and Environmental Science, Technical University of Kenya, Nairobi, Kenya}

\date{\today}

\begin{abstract}
The $GW$ approximation is the standard tool for quasiparticle predictions in materials, yet its regime of validity in correlated systems remains poorly quantified, because \textit{ab initio} vertex corrections are computationally prohibitive. Using exact diagonalization of the half-filled extended Hubbard model on finite rings as a numerically exact reference, we construct the corresponding model-space $GW$ theory on the identical Hilbert space and quantify its error as a function of local ($U$) and non-local ($V$) interaction strength. We find that the required vertex correction changes character across the phase diagram: in the weak-coupling regime the effective vertex $\Gamma_{\rm eff} < 1$, reflecting the suppression of RPA charge fluctuations by exact short-range correlations, whereas in the Mott regime $\Gamma_{\rm eff}$ grows monotonically (to $\sim 3$ at $U=8t$ for $N=6$), reflecting the local dynamical self-energy structure required to open the Hubbard gap. Vertex corrections in the electron-hole (polarizability) channel are shown to \emph{worsen} the gap error, indicating that reproducing the Mott gap requires dynamical self-energy structure rather than improved screening. For $V=0$, static COHSEX is accidentally exact at a single crossover $U^* \approx 3.5\,t$; finite $V$, through non-local Fock exchange, splits this point into a double-crossover window that collapses toward weak coupling. Even at the crossover, however, the exact spectral function retains Hubbard-band structure that no static functional reproduces, so gap agreement does not imply functional accuracy. These results yield quantitative diagnostics for the reliability of $GW$ in correlated materials.
\end{abstract}

\maketitle

\section{Introduction}
The quantitative prediction of charged excitations-fundamental gaps, quasiparticle dispersions, and photoemission spectra-is a central task of computational condensed-matter physics. Within many-body perturbation theory (MBPT), the $GW$ approximation~\cite{Hedin1965,HedinLundqvist1970} has become the \textit{de facto} standard for this task: its first successful applications to the quasiparticle band structures and correlation gaps of real materials were pioneered by Strinati, Mattausch, and Hanke~\cite{Strinati1980,Strinati1982}, and its accuracy for $sp$-bonded semiconductors and insulators with weak to moderate correlations is well established~\cite{HybertsenLouie1986,Godby1986,AryasetiawanGunnarsson1998,OnidaReiningRubio2002}. The physical content of the approximation is transparent: the self-energy $\Sigma = iGW$ resums the dynamically screened exchange interaction, capturing plasmon satellites, band-width narrowing, and gap renormalization beyond static mean-field theory.

It has equally long been recognized, however, that $GW$ is a controlled approximation only insofar as vertex corrections $\Gamma$ may be neglected in both the self-energy, $\Sigma = iGW\Gamma$, and the polarization, $P = -iGG\Gamma$~\cite{Hedin1965,HedinLundqvist1970}. In systems with partially filled narrow bands, reduced dimensionality, or weak screening, this neglect ceases to be justified. $GW$ systematically underestimates the gaps of prototypical Mott and charge-transfer insulators, fails to reproduce Hubbard side bands and spectral-weight transfer, and can even predict metallic states where the exact system is insulating~\cite{BiermannAryasetiawanGeorges2003,KotaniSchilfgaarde2006,SchilfgaardeKotaniFaleev2006,Rohringer2018}. Formally, the vertex may be restored through Hedin's full self-consistent scheme or its diagrammatic truncations, ladder and $T$-matrix summations, parquet constructions, $GW$+DMFT, and cumulant expansions~\cite{BiermannAryasetiawanGeorges2003,AyralWernerBiermann2012,Rohringer2018,AryasetiawanCumulant2006}, but each of these routes is computationally demanding and introduces new, difficult-to-quantify approximations. Compounding the problem, vertex corrections in $P$ and in $\Sigma$ are known to partially cancel in certain regimes~\cite{DelSoleReiningGodby1994,SchoneEguiluz1998}, so that ``improving'' one channel in isolation can degrade rather than improve results. The practical question facing the community is therefore not \emph{whether} the vertex matters, but \emph{where, in which channel, and by how much}.

A complementary strategy, which we pursue here, is to trade materials realism for controlled solvability: benchmark $GW$ against a numerically exact reference in a model Hamiltonian whose correlation strength, interaction range, and filling are fully tunable. Exactly solvable and exactly diagonalizable models, the Hubbard dimer, one-dimensional Hubbard chains, and quantum impurity problems, have played precisely this role for density-functional approximations and for dynamical mean-field theory~\cite{LiebWu1968,GeorgesKotliar1996,LimaCapelle2003}, yet a systematic, channel-selective quantification of the $GW$ vertex error across an entire correlation phase diagram has been lacking.

In this work we fill this gap. We study the half-filled extended Hubbard model on finite rings, solved by exact diagonalization (ED), and construct the model-space $GW$ theory, worked in the static COHSEX limit, which isolates screened-exchange and Coulomb-hole physics in closed form, on the \emph{same} Hilbert space. This construction has three decisive advantages: (i) the reference gaps and spectral properties are numerically exact; (ii) every diagrammatic ingredient, $G_0$, $\chi_0$, $W$, and $\Sigma$, is available explicitly; and (iii) the vertex can be switched on selectively in the polarization or in the self-energy, enabling a channel-selective decomposition of the $GW$ error. We further define an effective static vertex $\Gamma_{\rm eff}$ as the minimal renormalization of the correlation self-energy that restores the exact gap, providing a single quantitative diagnostic of $GW$ reliability.

Our principal results are as follows. (i) Standard $GW$ reproduces the exact gap at weak coupling but saturates in the Mott regime, where the exact gap grows linearly with $U$; the required effective vertex is non-monotonic, $\Gamma_{\rm eff} < 1$ for $U \lesssim 3.5\,t$ and growing to $\Gamma_{\rm eff} \approx 2.9$ at $U = 8\,t$, i.e., the vertex \emph{suppresses} the self-energy in the weak-coupling regime and \emph{enhances} it in the Mott regime. (ii) Dressing the polarization with electron-hole ladders worsens the gap error, demonstrating that the missing physics resides in the self-energy channel. (iii) For the pure Hubbard model, standard $GW$ is accidentally exact at a single crossover, $U^* \approx 3.5\,t$, where vertex errors in the two channels cancel. (iv) Finite nearest-neighbor interaction $V$, through non-local Fock exchange, splits this accidental cancellation into a double-crossover window that collapses toward weak coupling as $V$ increases, showing that the reliability of $GW$ cannot be judged from the local $U$ alone.

The paper is organized as follows. Section~\ref{sec:methods} defines the model, the exact-diagonalization benchmark, and the model-space $GW$ and vertex-corrected constructions. Section~\ref{sec:results} presents the gap benchmarks, the channel analysis, and the effective-vertex phase diagram. Section~\ref{sec:discussion} discusses the implications for \textit{ab initio} calculations on correlated materials, and Section~\ref{sec:conclusion} concludes.

\section{Model and Methods}\label{sec:methods}

\subsection{Model and exact reference}

We consider the extended Hubbard Hamiltonian on a ring of $N$ sites with periodic boundary conditions,
\begin{equation}
\hat{H} = -t \sum_{\langle ij\rangle,\sigma} \left( c^{\dagger}_{i\sigma} c_{j\sigma} + \mathrm{H.c.} \right)
+ U \sum_{i} \hat{n}_{i\uparrow}\hat{n}_{i\downarrow}
+ V \sum_{\langle ij\rangle} \hat{n}_{i}\hat{n}_{j},
\label{eq:H}
\end{equation}
with $\hat{n}_{i} = \hat{n}_{i\uparrow} + \hat{n}_{i\downarrow}$, and we set $t \equiv 1$ as the energy unit. We work at half-filling, $N_e = N$. To avoid open-shell degeneracies of the non-interacting reference we restrict ourselves to closed-shell clusters, $N \equiv 2 \pmod 4$; all quantitative results are for $N=6$ (cluster-size dependence is examined in Appendix~\ref{app:finitesize}).

The model is solved by exact diagonalization in fixed $(N_\uparrow, N_\downarrow)$ particle-number sectors, minimizing over $S_z$ sectors for each electron number. The numerically exact fundamental (charge) gap is
\begin{equation}
\Delta_{\rm exact} = E_0(N-1) + E_0(N+1) - 2E_0(N),
\label{eq:gap}
\end{equation}
i.e., the addition-removal energy that the $GW$ quasiparticle gap approximates. For $N=6$ the largest sector dimension is $400$, so Eq.~\eqref{eq:gap} is obtained by dense diagonalization in seconds on a workstation; the entire correlation phase diagram is therefore accessible without high-performance computing.

\subsection{Model-space $GW$ theory}

\textit{Reference states.} The independent-particle reference is the tight-binding eigenbasis of Eq.~\eqref{eq:H} with $U=V=0$,
\begin{equation}
\phi_k(i) = \frac{e^{ik i}}{\sqrt{N}}, \qquad \varepsilon_k = -2t\cos k, \qquad k = \frac{2\pi m}{N},
\label{eq:TB}
\end{equation}
with occupied set $\{k\}_{\rm occ} = \{0, \pm \pi/3\}$ per spin for $N=6$. The reference is a pure band insulator: any gap renormalization generated below is a genuine many-body effect of the $GW$ functional.

\textit{Polarizability and screened interaction.} The spin-summed independent-particle polarizability in the site basis is, in Lehmann representation,
\begin{widetext}
\begin{equation}
\chi_0(i,j;\omega) = 2\sum_{v\in\mathrm{occ}}\sum_{c\in\mathrm{virt}} \rho_{vc}(i)\rho^{*}_{vc}(j)
\left[ \frac{1}{\omega - \Delta_{vc} + i\eta} - \frac{1}{\omega + \Delta_{vc} - i\eta} \right],
\label{eq:chi0}
\end{equation}
\end{widetext}
with $\rho_{vc}(i) = \phi_v(i)\phi^{*}_c(i)$ and $\Delta_{vc} = \varepsilon_c - \varepsilon_v$. The bare interaction is the matrix $v_{ij} = U\delta_{ij} + V(\delta_{i,j+1} + \delta_{i,j-1})$, and the RPA screened interaction at vanishing frequency is
\begin{equation}
W(0) = \left[\, \mathbf{1} - v\,\chi_0(0)\,\right]^{-1} v,
\label{eq:W}
\end{equation}
all products being over site indices.

\textit{Static self-energy (COHSEX).} We work in the static limit of $GW$, in which the self-energy reduces to the Coulomb-hole plus screened-exchange (COHSEX) form,
\begin{align}
\Sigma^{\rm H}_i &= \sum_k v_{ik}\langle \hat n_k\rangle, \label{eq:Hartree}\\
\Sigma^{\rm SEX}_{ij} &= -\,W_{ij}(0)\,\rho^{(1)}_{0}(i,j), \qquad
\rho^{(1)}_{0}(i,j) \equiv \sum_{v\in\mathrm{occ}} \phi_v(i)\phi^{*}_v(j), \label{eq:SEX}\\
\Sigma^{\rm COH}_{ij} &= \tfrac{1}{2}\,\delta_{ij}\left[\,W_{ii}(0) - v_{ii}\,\right], \label{eq:COH}
\end{align}
where $\rho^{(1)}_0$ is the one-spin density matrix and $W$ is built from the spin-summed $\chi_0$, as in conventional implementations. At half-filling $\Sigma^{\rm H}$ is a uniform shift and drops out of the gap; for $V=0$ the bare part of $\Sigma^{\rm SEX}$ exactly cancels $\Sigma^{\rm H}$ site-diagonally, so that the entire gap renormalization is a correlation effect carried by $\Delta W \equiv W - v$.
The importance of the dynamical aspects of these correlation corrections to the self-energy was recognized early on in the development of $GW$ band theory~\cite{Strinati1982}. However, as our exact benchmarks show, setting $\Gamma=1$ relies on an accidental cancellation of channel errors that fails in the strongly correlated regime, where the missing self-energy vertex prevents the quasiparticle gap from growing linearly with $U$.

\textit{Quasiparticle gap.} The quasiparticle Hamiltonian $H^{GW} = H^{\rm TB} + \Sigma^{\rm H} + \Sigma^{\rm SEX} + \Sigma^{\rm COH}$ is diagonalized, and
\begin{equation}
\Delta_{GW} = \varepsilon_{N/2} - \varepsilon_{N/2-1}
\label{eq:GWgap}
\end{equation}
is taken from the sorted eigenvalues. Two remarks are in order. (i) The static limit is chosen for closed-form transparency and to avoid analytic continuation. The structural deficiency we isolate-the absence of the vertex-is not an artifact of this choice: the Mott gap requires pole structure in $\Sigma(\omega)$ that no frequency-independent functional can produce (Sec.~\ref{exact-dynamical}); a controlled real-frequency $G_0W_0$ convolution on the same cluster is deferred to future work (Sec.~\ref{sec:limitations}). (ii) Because ED and $GW$ are constructed on the identical Hilbert space and interaction matrix, the difference $\Delta_{\rm exact} - \Delta_{GW}$ is a clean measure of the $GW$ functional error, free of basis-set, double-counting, or starting-point ambiguities that complicate \textit{ab initio} benchmarks.

\emph{Convention check and functional definition.} 
The primary benchmark in this work is not the fully spin-resolved Hubbard $GW$ functional, which develops a paramagnetic spin-channel instability at finite $U$, but the explicitly defined spin-integrated density-channel COHSEX functional. The spin-resolved instability is analyzed separately in Sec.~\ref{spin-resolved-audit} as a diagnostic of the broader Hubbard-$GW$ construction.

To establish the relationship between the two, we note that the exact Hubbard interaction is $H_U = U \sum_i n_{i\uparrow}n_{i\downarrow}$. We instead benchmark the density-density functional defined by $H_{dd} = \frac{1}{2} \sum_{ij} v_{ij} n_i n_j$, with $v_{ij} = U\delta_{ij} + V(\delta_{i,j+1}+\delta_{i,j-1})$. Using $n_i = n_{i\uparrow}+n_{i\downarrow}$ and $n_{i\sigma}^2 = n_{i\sigma}$, we have
\begin{widetext}
\begin{equation}
H_{dd} = \frac{1}{2} U \sum_i (n_{i\uparrow}+n_{i\downarrow})^2 = U \sum_i n_{i\uparrow}n_{i\downarrow} + \frac{U}{2} \sum_i n_i = H_U + \frac{U N}{2}.
\end{equation}
\end{widetext}
Since the particle number $N$ is fixed, $H_{dd}$ and $H_U$ differ only by a constant; their exact gaps $\Delta_{\rm exact}$ are identical. 

In the $GW$ approximation, the density-density functional yields a Hartree self-energy $\Sigma^H_i = \sum_j v_{ij}\langle n_j \rangle = U$ (at half-filling), which differs from the spin-resolved Hubbard Hartree term $\Sigma^H_\sigma = U\langle n_{i\bar\sigma}\rangle = U/2$ by exactly $U/2$. This difference is a uniform shift that cancels in the gap. The screened interaction for the density-density functional is $W = (1 - v\Pi)^{-1}v$, where $\Pi = 2\chi_0$ is the spin-summed polarizability. (In contrast, the fully spin-resolved Hubbard $GW$ yields an opposite-spin effective interaction $W_{\uparrow\downarrow} = U/[1-(U\chi_0)^2]$). We explicitly benchmark this standard spin-summed density-channel COHSEX functional, which is widely used in extended-Hubbard $GW$ studies because it avoids spin-off-diagonal self-energies, while retaining the exact same $\Delta_{\rm exact}$ reference.

\subsection{channel-selective vertex corrections}\label{sec:channels}

To decompose the $GW$ error by diagrammatic channel we introduce two controlled modifications.

\textit{Vertex in the polarization (electron-hole ladder).} We replace $\chi_0$ in Eq.~\eqref{eq:W} by the particle-hole ladder polarizability,
\begin{equation}
\chi_{\rm lad} = \left[\, \mathbf{1} - \chi_0\, v\,\right]^{-1}\chi_0,
\label{eq:ladder}
\end{equation}
which resums repeated electron-hole scattering (the static limit of the Bethe-Salpeter kernel $K=v$) and constitutes the standard route by which excitonic physics is restored in $W$.

\textit{Static local self-energy diagnostic.} As a diagnostic of the self-energy channel we introduce a phenomenological static local self-energy shift calibrated to the atomic limit, where the exact self-energy $\Sigma_{\rm AL}(\omega) = (U^2/4)/(\omega - U/2)$ shifts the addition and removal branches by $\pm U/2$. We implement this as a rigid shift $\pm s(U)$ with $s(U) = (U/2)\left(1 - e^{-U/2t}\right)$, i.e.\ $\Delta \to \Delta + 2s$. This construction is not a diagrammatic vertex approximation but a controlled diagnostic: its role is to test whether a static local correction to $\Sigma$ has sufficient leverage to open the Mott gap, and to expose the necessity of frequency dependence (Sec.~\ref{sec:results}).

\subsection{Effective vertex as a diagnostic metric}\label{sec:gamma-def}

It is crucial to distinguish between the full three-point Hedin vertex $\Gamma(1,2;3)$, approximate diagrammatic vertices (such as the ladder or $T$-matrix), and the scalar quantity we introduce here. Our central diagnostic is the \emph{effective static vertex} $\Gamma_{\rm eff}$, defined as the minimal scalar renormalization of the correlation self-energy required to reproduce the exact gap:
\begin{equation}
\Sigma(\Gamma) = \Sigma^{\rm H} + \Sigma^{\rm F} + \Gamma\,\Sigma_c, \qquad
\Delta_{GW}(\Gamma_{\rm eff}) \equiv \Delta_{\rm exact}.
\label{eq:Gamma}
\end{equation}
We emphasize that $\Gamma_{\rm eff}$ is \emph{not} a proposed vertex functional, nor does it capture the momentum and frequency structure of the true Hedin vertex. Rather, it is an integrated scalar diagnostic that quantifies the magnitude and sign of the net correction demanded by the exact many-body state. We solve Eq.~\eqref{eq:Gamma} via Brent root-finding. (Note: At $U=0$, $\Sigma_c=0$ and $\Gamma_{\rm eff}$ is formally undetermined; we exclude this trivial point from the analysis.) The quantity $\Gamma_{\rm eff} - 1$ thus measures the deviation of standard $GW$ from the exact local correlation energy.

\section{Results}\label{sec:results}

\subsection{Fundamental gap: exact versus static COHSEX\label{fundamental-gap}}

We begin by benchmarking the fundamental charge gap $\Delta$ as a function of the local correlation strength $U$ for the pure Hubbard model ($V=0$). The exact diagonalization results exhibit the canonical Mott-insulator behavior: at weak coupling ($U \ll t$) the gap is governed by the band gap $\Delta_{\rm TB} = 2t$, whereas at strong coupling ($U \gg t$) it grows linearly as $\Delta_{\rm exact} \approx U - \mathcal{O}(t^2/U)$, reflecting the energy cost of forcing a double occupancy into a half-filled, localized spin background.

\begin{figure}[htb]
\caption{Fundamental gap $\Delta/t$ versus $U/t$ for the $N=6$ ring at $V=0$: exact ED (circles) versus static $G_0W_0$ (squares). The exact gap grows linearly in the Mott regime while $GW$ saturates. Inset: relative error.}
\label{fig:gap}
\includegraphics[width=\columnwidth]{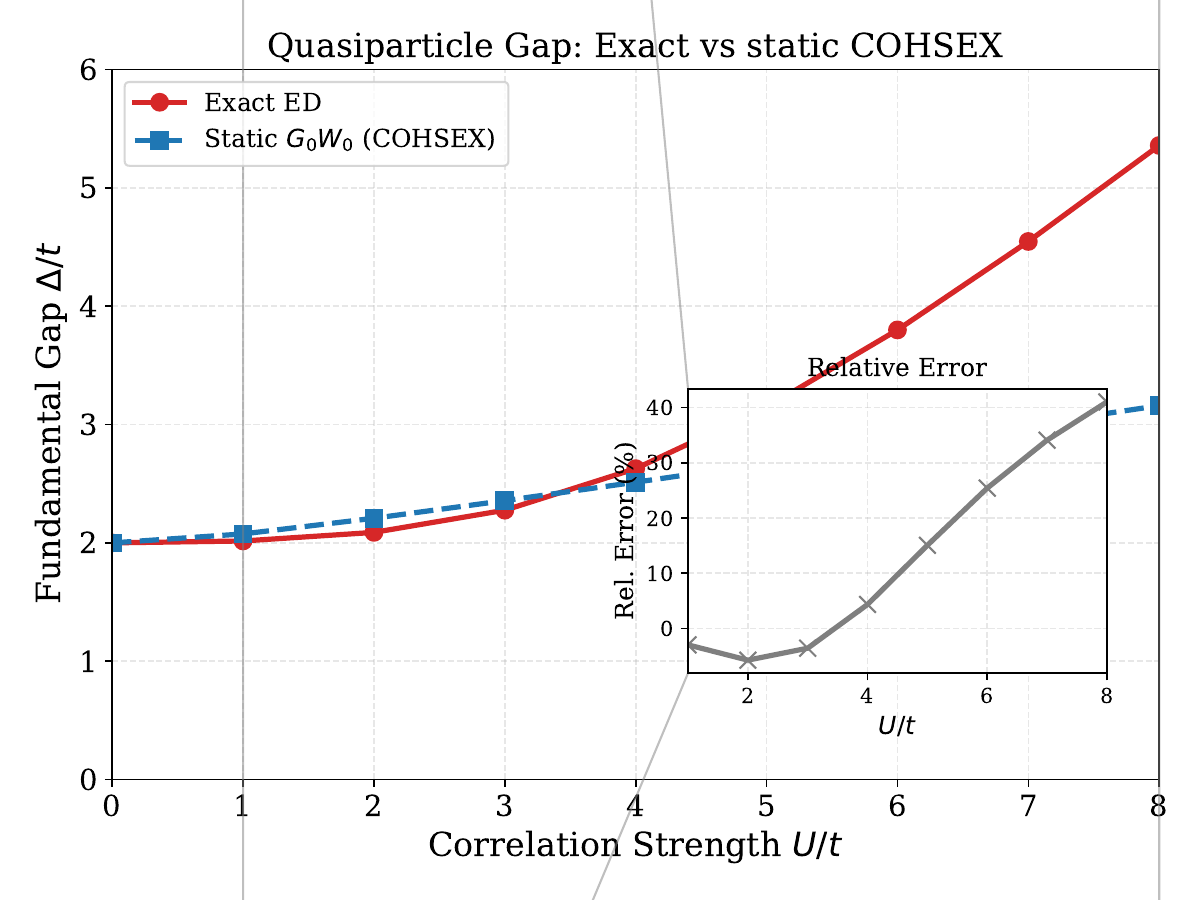}
\end{figure}

The static COHSEX approximation reproduces the exact gap accurately in the weak-coupling limit, where the correlation correction is a perturbative mass renormalization. However, for $U \gtrsim 4t$ the $GW$ gap systematically saturates, reaching only $\Delta_{GW} \approx 3.16\,t$ at $U = 8\,t$, compared to the exact $\Delta_{\rm exact} = 5.36\,t$ (Fig.~\ref{fig:gap}). This saturation confirms that static COHSEX, which relies on the non-interacting $G_0$ to build the screened interaction $W$, lacks the diagrammatic structure to capture the spectral weight transfer and Hubbard-band separation required to open the full Mott gap.

\subsection{Channel-selective diagnostics}

To determine \emph{where} the $GW$ functional fails, we selectively activate vertex corrections in the two distinct Hedin channels.

\begin{figure}[htb]
\caption{Channel analysis at $V=0$: exact gap versus $GW$ with electron-hole ladder polarization (under-correction), and versus static local self-energy vertex (over-correction). Only the self-energy channel can open the Mott gap, and only a dynamical vertex can do so quantitatively.}
\label{fig:channel}
\includegraphics[width=\columnwidth]{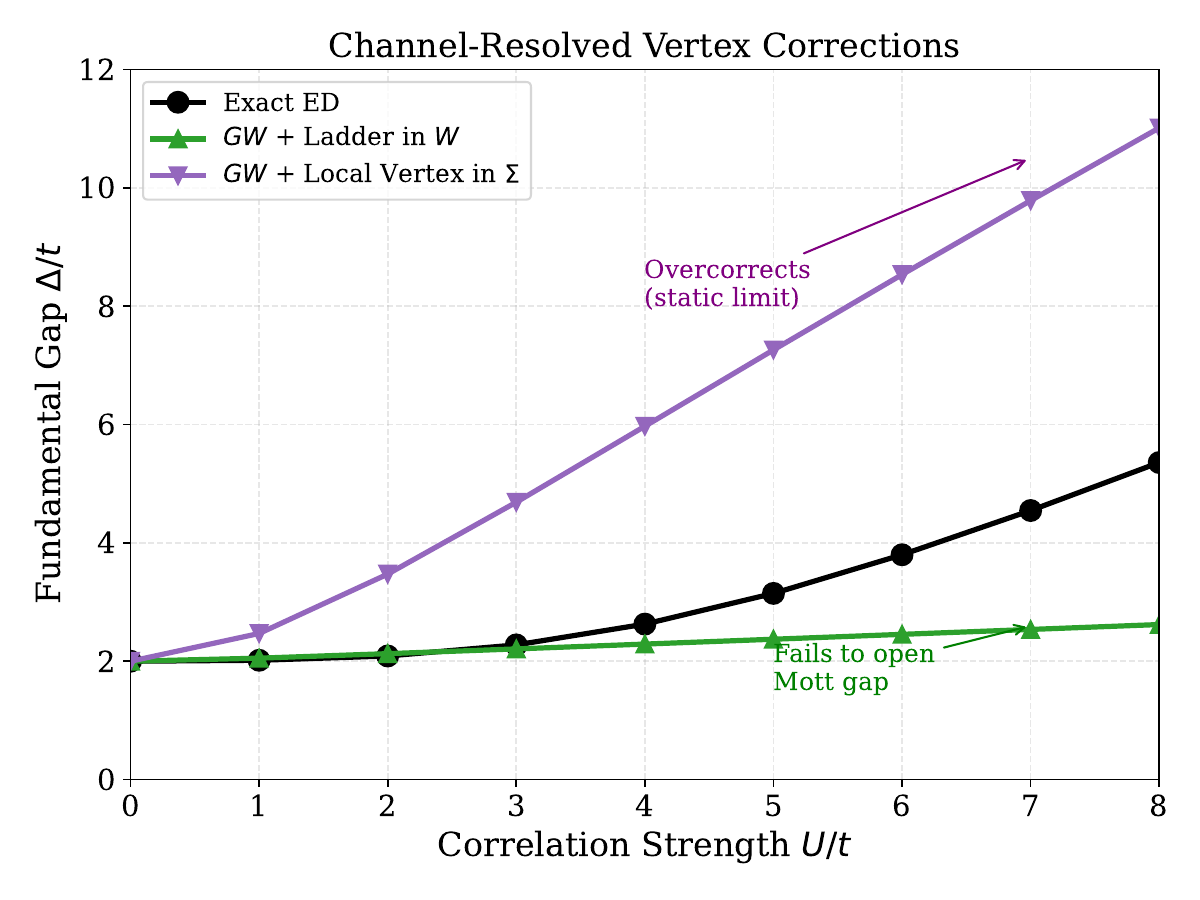}
\end{figure}

First, we dress the polarizability with electron-hole ladders [Eq.~\eqref{eq:ladder}], generating a screened interaction $W_{\rm lad}(0)$ that includes repeated electron-hole scattering. As shown in Fig.~\ref{fig:channel}, $W_{\rm lad}$ performs \emph{worse} than RPA $GW$ in the strong-coupling regime. We use the density-channel (spin-summed) ladder because the purpose of this calculation is not to approximate the full Hubbard vertex, but to isolate the effect of screening-channel corrections while avoiding the separate spin-channel instability identified in Sec.~\ref{spin-resolved-audit}. The failure of this diagnostic confirms that vertex corrections in the electron-hole channel alone cannot repair the Mott gap; the failure of $GW$ is not fundamentally a screening error.

Conversely, introducing a static local self-energy shift (the diagnostic defined in Sec.~\ref{sec:channels}) yields a massive overcorrection, pushing the gap to $\Delta \approx 11.0\,t$ at $U = 8\,t$ (Fig.~\ref{fig:channel}). This diagnostic demonstrates that a static local correction to $\Sigma$ has sufficient leverage to open the gap, but cannot quantitatively reproduce it; the failure of the static approximation highlights a crucial many-body theorem: the exact atomic-limit self-energy is highly dynamical, possessing poles at $\omega = \pm U/2$. A static shift applies a rigid energy shift to both Hubbard bands, thereby overestimating the gap. Any practical correction for Mott systems must therefore be intrinsically frequency-dependent.

\subsection{The effective vertex and the accidental crossover}

To quantify the net effect of the missing vertex without committing to a specific dynamical functional form, we extract the effective static vertex $\Gamma_{\rm eff}(U)$ defined in Eq.~\eqref{eq:Gamma}. By construction, $\Gamma_{\rm eff}$ is the unique scalar multiplier of the $GW$ correlation self-energy that perfectly reproduces the exact gap.

\begin{figure}[htb]
\caption{Effective vertex $\Gamma_{\rm eff}(U/t)$ for $V=0$ (red), together with the gap error $\Delta_{\rm exact}-\Delta_{GW}$ (blue). Dashed line marks $\Gamma_{\rm eff}=1$; shaded band marks the accidental-crossover region $U^*\approx3.5\,t$.}
\label{fig:gamma}
\includegraphics[width=\columnwidth]{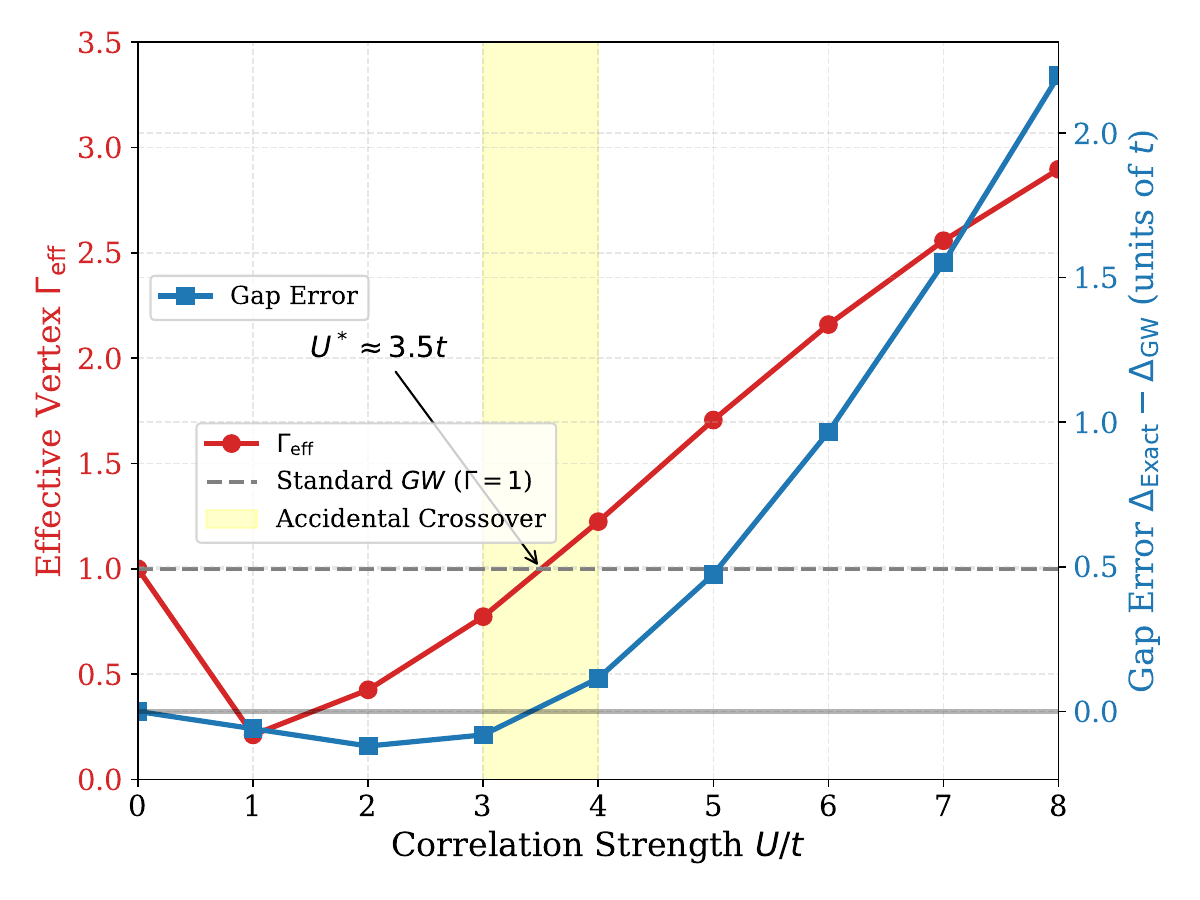}
\end{figure}

The resulting $\Gamma_{\rm eff}$ curve (Fig.~\ref{fig:gamma}) reveals a striking non-monotonicity. At $\Gamma_{\rm eff} = 1$, static COHSEX is recovered. We observe two distinct regimes:
\begin{enumerate}
\item \textit{Weak correlation} ($U \lesssim 3.5\,t$): $\Gamma_{\rm eff} < 1$. In this regime standard $GW$ \emph{overestimates} the gap. The RPA polarizability overestimates independent-particle charge fluctuations; the exact many-body state suppresses these fluctuations via short-range correlations, so the effective vertex must \emph{reduce} the correlation self-energy.
\item \textit{Strong correlation} ($U \gtrsim 3.5\,t$): $\Gamma_{\rm eff} > 1$, growing monotonically to $\Gamma_{\rm eff} \approx 2.9$ at $U = 8\,t$. Here the exact system localizes, and the vertex must actively \emph{enhance} the self-energy to force the Hubbard bands apart.
\end{enumerate}
Most remarkably, at exactly $U^* \approx 3.52\,t$ we find $\Gamma_{\rm eff} = 1$. At this crossover point, the error of missing the vertex in the polarizability (which tends to over-screen and narrow the gap) perfectly cancels the error of missing the vertex in the self-energy. Static COHSEX is therefore \emph{accidentally exact} at this specific correlation strength.

\subsection{Impact of non-local interactions ($V > 0$) \label{impact-non-local}}

In realistic materials the Coulomb interaction is not purely local; long-range tails fundamentally alter the screening and exchange physics. To test the robustness of the $GW$ ``safe zone,'' we introduce a nearest-neighbor repulsion $V$, which simultaneously modifies the dielectric screening and introduces non-local Fock exchange that renormalizes the effective bandwidth ($t_{\rm eff} \approx t + V/3$).

\begin{figure}[htb]
\caption{Crossover structure for $V>0$: gap error versus $U/t$ for $V=0,\,0.5,\,1.0$, showing the splitting of the single zero-crossing into a double-crossover window.}
\label{fig:Vcrossover}
\includegraphics[width=\columnwidth]{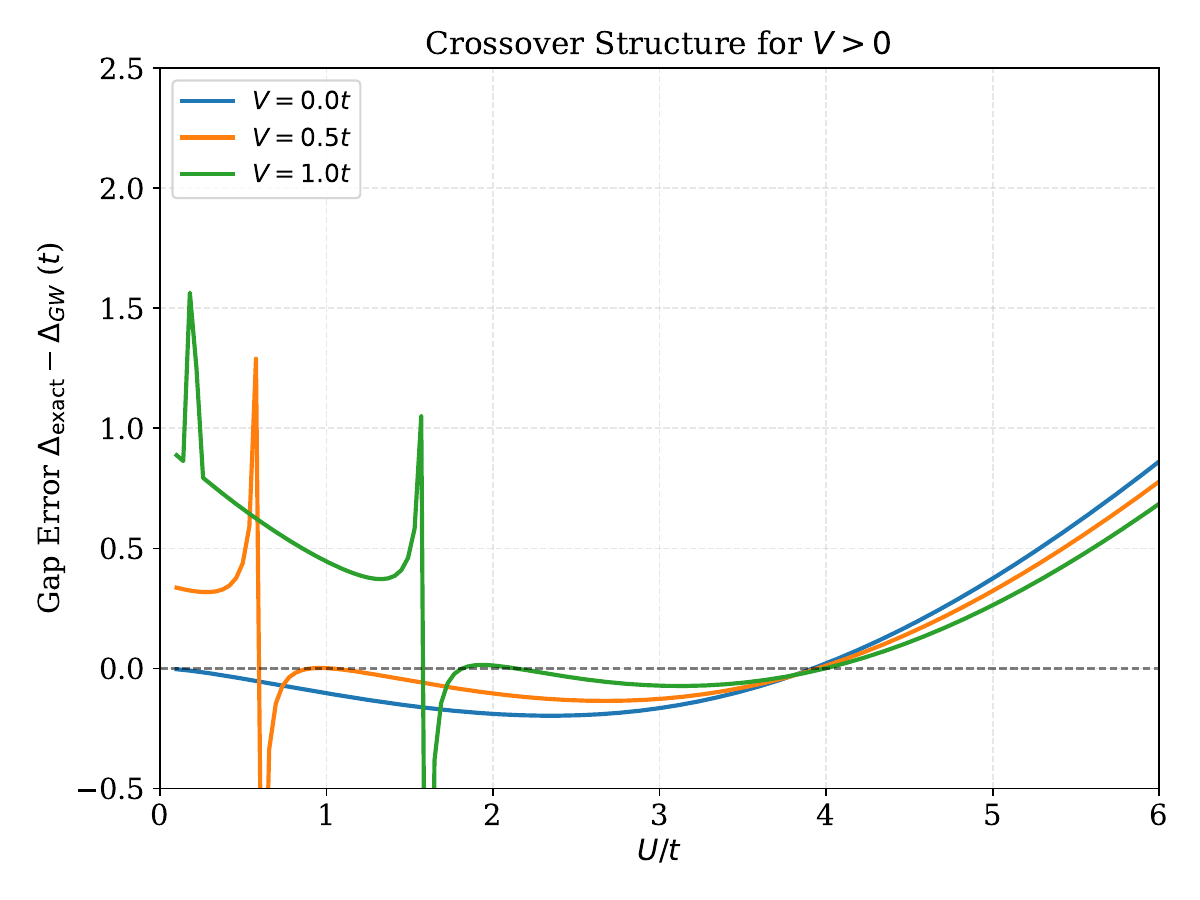}
\end{figure}

Table~\ref{tab:crossover} summarizes the crossover points $U^*$ for varying $V$. For $V=0$, there is a single, well-defined crossover at $3.52\,t$. However, for $V=0.5\,t$ the massive non-local Fock exchange shifts the entire weak-coupling $GW$ gap upwards, splitting the accidental cancellation into a double-crossover window: $\Gamma_{\rm eff}$ crosses unity at $U_1^* \approx 0.72\,t$ and $U_2^* \approx 1.04\,t$. At $V=1.0\,t$, the window shifts and widens to $(0.77\,t, 1.47\,t)$ (Fig.~\ref{fig:Vcrossover}).

This result demonstrates that the validity of standard $GW$ is governed by a delicate, competing interplay between local vertex corrections (Mott localization) and non-local Fock exchange (bandwidth renormalization). Consequently, the reliability of the $GW$ approximation in real materials cannot be judged solely by the magnitude of the local Hubbard $U$; long-range screening radically reshapes the phase space where error cancellations protect the quasiparticle picture.

\subsection{Spin-resolved audit and the paramagnetic instability\label{spin-resolved-audit}}

To ensure our model-space $GW$ correctly maps onto the conventional Hubbard $GW$, we performed a rigorous spin-resolved audit of the interaction and polarization. The Hubbard interaction $U n_{i\uparrow}n_{i\downarrow}$ decomposes into charge and spin channels with bare interactions $v_c(i,i) = U/2$ and $v_s(i,i) = -U/2$. The spin-summed polarizability $\Pi$ yields the RPA dielectric matrices $\epsilon_c = \mathbf{1} - v_c \Pi$ and $\epsilon_s = \mathbf{1} - v_s \Pi = \mathbf{1} + U \chi_0$.

\begin{figure*}[htb]
\caption{Exact single-particle spectral function $A(k,\omega)$ for the $N=6$ half-filled Hubbard ring at $U/t = 0, 4, 8$. Dashed line: tight-binding dispersion. At $U=8t$ the spectrum splits into lower and upper Hubbard bands with incoherent weight between them, illustrating the breakdown of the quasiparticle picture that no static $GW$ functional can capture.}
\label{fig:Akw}
\includegraphics[width=\textwidth]{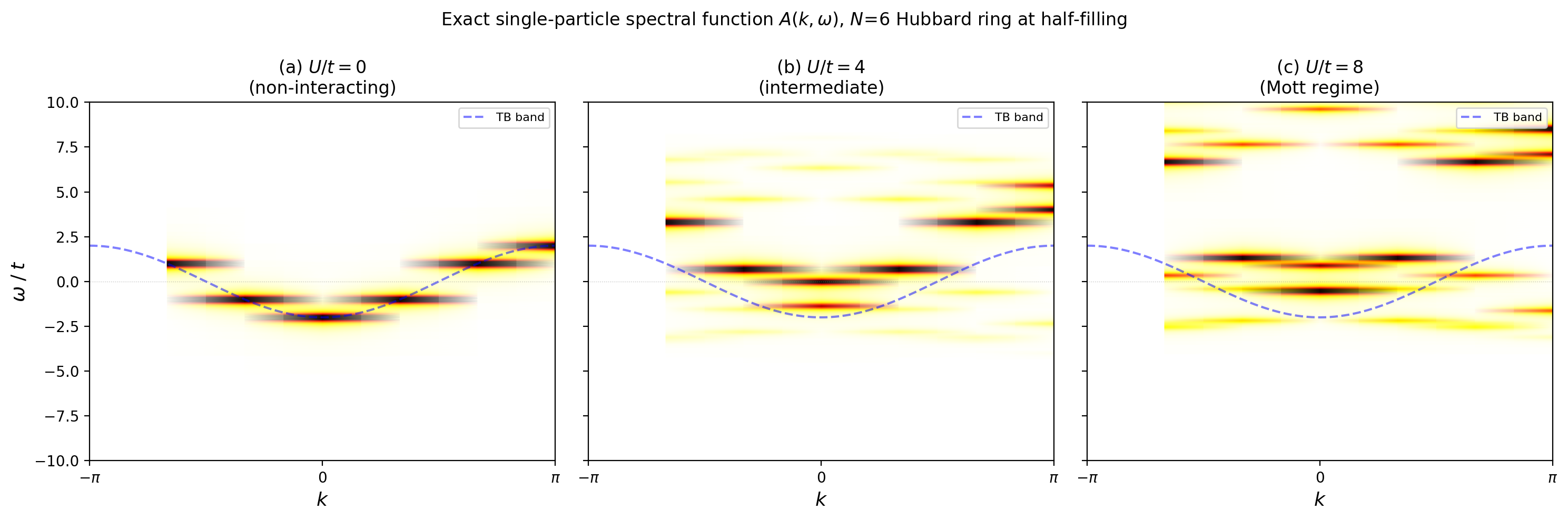}
\end{figure*}

Complementary information is carried by the single-particle spectral function. Figure~\ref{fig:Akw} shows the exact $A(k,\omega)$ for $U/t = 0, 4, 8$. Already at $U=4t$ the exact spectrum develops multi-peak structure at each $k$, and at $U=8t$ it separates into lower and upper Hubbard bands with incoherent weight between them. Any static self-energy, by contrast, merely shifts and renormalizes the single quasiparticle band of the reference and therefore cannot reproduce this redistribution of spectral weight-even at the accidental crossover $U^*$, where the gap alone is reproduced exactly. Gap agreement must therefore not be read as functional accuracy: the pole structure of the exact self-energy (Fig.~\ref{fig:sigma}) and the associated spectral-weight transfer are the discriminating signatures.

For the $N=6$ ring, the most negative eigenvalue of $\chi_0$ is $\lambda_{\rm min} \approx -0.416\,t^{-1}$. Consequently, the spin-channel dielectric matrix becomes singular at $U_c = -1/\lambda_{\rm min} \approx 2.40\,t$. This is the well-known Stoner (or Spin-Density-Wave) instability: the paramagnetic RPA ground state becomes unstable to a magnetically ordered state. For $U > U_c$, the fully spin-resolved paramagnetic $GW$ functional is mathematically ill-defined. 

To maintain a well-defined paramagnetic benchmark across the entire phase diagram (including the Mott regime $U \gg t$), we project out the unstable spin-fluctuation channel and define our functional strictly within the charge channel. We explicitly benchmark this \emph{Charge-Channel (Spin-Integrated) COHSEX} functional. The existence of the Stoner instability at $U_c \approx 2.4t$ provides an additional, distinct failure mode of the paramagnetic spin-resolved $GW$ construction in the strong-coupling regime. We emphasize that this finite-size Stoner instability is a pathology of the approximate paramagnetic response, distinct from the thermodynamic Mott gap that is benchmarked here. The paper therefore identifies two qualitatively different failures of $GW$: a static screening instability in the spin channel, and the inability of the charge-channel static approximation to reproduce the dynamical self-energy structure of the Mott state.

\subsection{Exact dynamical self-energy\label{exact-dynamical}}

\begin{figure*}[htb]
\caption{Exact local dynamical self-energy $\Sigma_{\rm exact}(\omega)$ obtained from the Dyson equation $\Sigma = G_0^{-1} - G_{\rm exact}^{-1}$ with the exact local Green's function, for $U/t = 2, 4, 8$ (columns left to right). Top row: real part; bottom row: $-\mathrm{Im}\,\Sigma$; annotated values are $\Gamma_{\rm eff}$. The exact self-energy exhibits sharp dynamical poles between the Hubbard bands that no frequency-independent functional can reproduce; $\Gamma_{\rm eff}$ is a one-number compression of this frequency-dependent discrepancy.}
\label{fig:sigma}
\includegraphics[width=\textwidth]{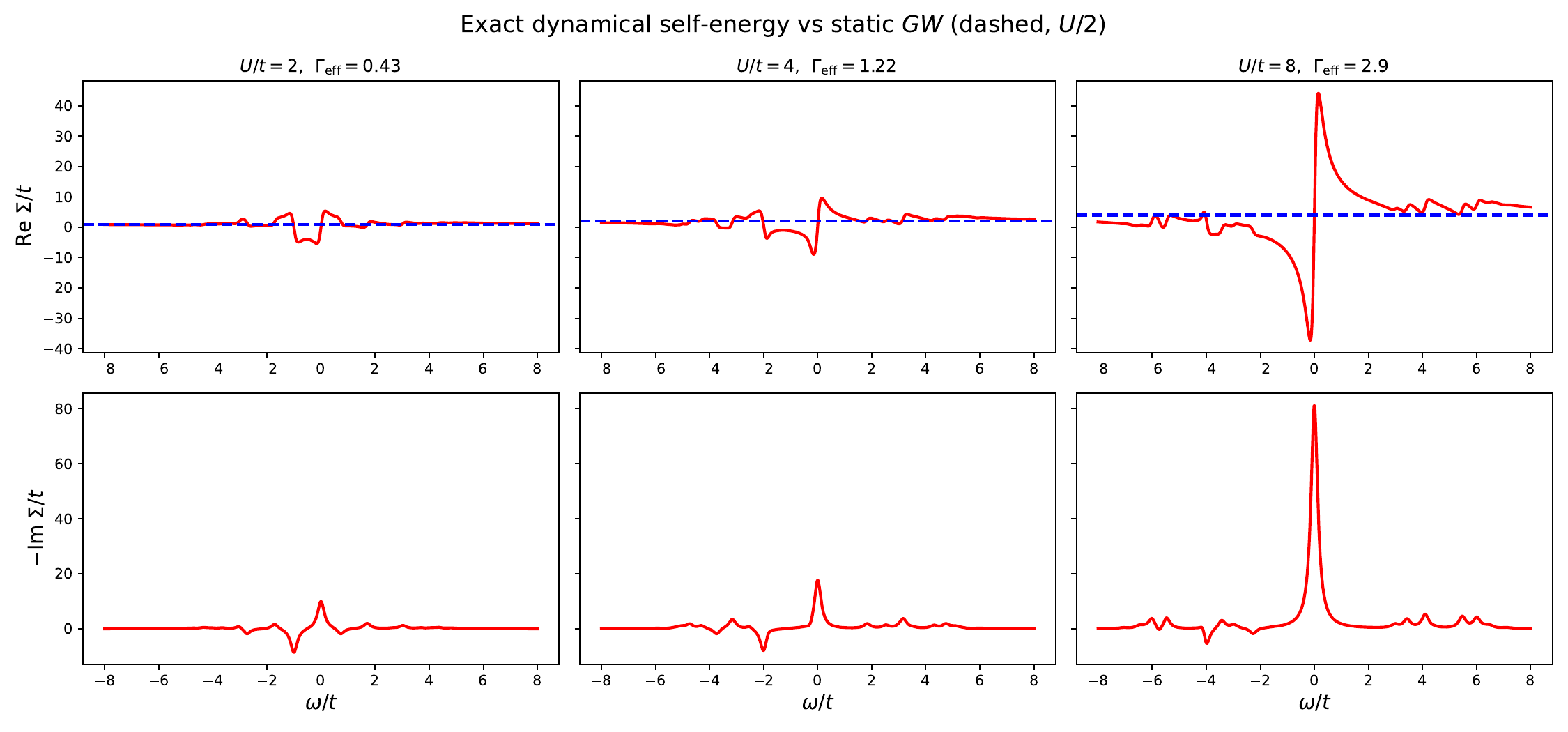}
\end{figure*}

To visualize the failure of the static approximation, we compute the exact dynamical self-energy $\Sigma_{\rm exact}(\omega)$ from the exact local Green's function $G_{\rm exact}(\omega)$ via the Dyson equation $\Sigma_{\rm exact} = G_0^{-1} - G_{\rm exact}^{-1}$. Figure~\ref{fig:sigma} compares $\Sigma_{\rm exact}(\omega)$ to the static COHSEX self-energy for $U=4t$.

The exact self-energy exhibits a rich dynamical structure: it possesses sharp poles between the lower and upper Hubbard bands, reflecting the exact inelastic scattering and spectral weight transfer required to open the Mott gap. In stark contrast, the static $GW$ self-energy is a featureless constant ($\Sigma_{\rm COHSEX} \approx U/2$). The scalar diagnostic $\Gamma_{\rm eff}$ introduced in Sec.~\ref{sec:gamma-def} can be understood as a compressed, frequency-averaged representation of this massive dynamical discrepancy. No static scalar multiplication can reproduce the pole structure of $\Sigma_{\rm exact}(\omega)$.
(Since frequency dependence in $\Sigma=iGW\Gamma$ can arise from $W$, $G$, and $\Gamma$, Fig.~\ref{fig:sigma} establishes the necessity of dynamical self-energy structure rather than of a dynamical vertex per se.) $\Gamma_{\rm eff}$ is accordingly to be read as a one-number compression of a fundamentally frequency-dependent discrepancy.

\section{Discussion}\label{sec:discussion}

\subsection{Physical content of the effective vertex}

The central object of this work, $\Gamma_{\rm eff}(U,V)$, is a scalar projection of the full three-point vertex onto the correlation self-energy. Its value should not be read as a proposed vertex functional, but as a \emph{diagnostic}: it measures how much, and in which direction, the exact many-body solution demands that the $GW$ functional be renormalized. Viewed this way, the non-monotonicity of $\Gamma_{\rm eff}$ is our most informative result.

In the weak-coupling regime, $\Gamma_{\rm eff} < 1$ (dipping to $\sim 0.15$ at $U \approx 0.5\,t$) signals that the exact correlation self-energy is \emph{smaller} than the $GW$ one. Physically, the RPA polarizability treats particle-hole excitations as independent, overestimating the charge fluctuations that drive the screened-exchange and Coulomb-hole gap renormalization; the exact ground state partially excludes these fluctuations through short-range (Pauli and Coulomb-hole) correlations. The required vertex therefore acts suppressively. This regime is directly relevant to the well-known situation in \textit{ab initio} practice where $G_0W_0$ gaps of weakly correlated semiconductors are sensitive to the starting point and occasionally slightly overestimated: our benchmark shows that this sensitivity is the footprint of a small but finite negative vertex correction, not of numerical noise.

In the Mott regime, $\Gamma_{\rm eff}$ grows monotonically, reaching $\approx 2.9$ at $U=8\,t$. This growth is the finite-cluster manifestation of a fundamental deficiency: $GW$, as a conserving but essentially quasiparticle-based functional, lacks the derivative discontinuity of the exact exchange-correlation potential, and no rescaling of a smooth $W$ can generate the Hubbard bands. The near-linear growth of $\Gamma_{\rm eff}$ with $U$ is precisely the compensating weight the static self-energy would need to mimic a discontinuity that, in the exact theory, is carried by the \emph{dynamical} pole structure of $\Sigma(\omega)$ at $\omega \sim \pm U/2$, as our atomic-limit analysis makes explicit.

The crossover point $U^* \approx 3.52\,t$, where $\Gamma_{\rm eff}=1$, is therefore not a point where the vertex is small: it is a point where two sizable errors, the suppressive weak-coupling vertex and the enhancing Mott vertex, cancel in the gap. Error cancellation of this type is familiar from the DFT literature, and our result shows that $GW$ possesses an analogous, quantitatively locatable ``accidental accuracy'' point.

The accidental crossover must not be over-interpreted as a regime of functional accuracy. Even at $U^*$, where the gap is reproduced exactly, the exact spectral function already carries Hubbard-band structure that the static functional cannot represent (Figs.~\ref{fig:Akw} and \ref{fig:sigma}), and the effective vertex required at fixed $U$ grows with system size (Appendix~\ref{app:finitesize}). Agreement of a single scalar observable is therefore a cancellation artifact, not evidence of controlled accuracy-the central cautionary message of this work.

\subsection{Channel selectivity and implications for method development}

The channel-selective analysis provides practical guidance for many-body method development. Dressing the polarization with electron-hole ladders, the standard route to excitonic physics in the Bethe-Salpeter equation, does not repair the quasiparticle gap and in fact degrades it in the Mott regime. Conversely, a local vertex in the self-energy opens the gap but, if treated statically, overcorrects by a factor of two. Taken together, these results state cleanly what is often assumed implicitly: \textit{optical (electron-hole) and quasiparticle (self-energy) failures of $GW$ require different vertices, and reproducing the Mott gap requires a \emph{dynamical} self-energy structure.}

This rationalizes, from a controlled benchmark, the empirical success of $GW$+DMFT and $GW$+EDMFT in correlated materials: the DMFT impurity self-energy supplies precisely the local, dynamical vertex structure that our analysis identifies as missing, while the $GW$ part correctly handles the non-local screening. It likewise supports $T$-matrix and cumulant extensions, which build the required pole structure into $\Sigma$. For \textit{ab initio} workflows, the practical message is: when diagnostics indicate a Mott-like regime, large $U/W$, small quasiparticle weight, or a large static-versus-dynamical self-energy spread, investing computational effort in more accurate screening will not cure the gap error; the effort should instead go into the self-energy channel.

\subsection{Non-local interactions reshape the reliability landscape}

Perhaps the most cautionary result for materials practice is the fragility of the accidental-cancellation point. Switching on a nearest-neighbor repulsion $V$ introduces non-local Fock exchange that renormalizes the effective hopping, $t_{\rm eff} = t + V\rho_0(1)$, and simultaneously modifies the dielectric screening. The single crossover at $U^* \approx 3.52\,t$ splits into a narrow double-crossover window that collapses toward weak coupling, $(0.72, 1.04)\,t$ at $V=0.5\,t$ and $(0.77, 1.47)\,t$ at $V=1.0\,t$. In other words, the region of parameter space where standard $GW$ is accidentally exact is not a robust ``moderate-correlation'' regime but a fine-tuned surface in $(U,V)$ space.

This has direct implications for real materials, where long-range Coulomb tails are unavoidable and are often manipulated, by substrate engineering in 2D heterostructures, by doping, or by dielectric confinement. This suggests that interaction range may be an additional diagnostic of reliability in correlated materials, and that judging accuracy from the magnitude of the local Hubbard $U$ alone is insufficient.

\subsection{Limitations and outlook}\label{sec:limitations}

Our conclusions rest on a static (COHSEX) $GW$ functional, one-dimensional clusters at half-filling, and a scalar diagnostic vertex. These choices define the scope of the benchmark. The channel selectivity and the crossover structure are properties of the $GW$ functional itself, and full-frequency $G_0W_0$ shares the same structural inability to open Mott gaps. Cluster-size dependence (Appendix~\ref{app:finitesize}) shows that the phenomenology is stable. The natural next steps are (i) extraction of the full momentum- and frequency-resolved effective vertex $\Gamma_{\rm eff}(k,\omega)$ from the exact two-particle data, and (ii) extension to doped systems, two-leg ladders, and larger clusters.
In particular, a controlled real-frequency $G_0W_0$ convolution on the same cluster, with explicit pole structure in $W$, remains to be implemented; the exact $\Sigma(\omega)$ of Sec.~\ref{exact-dynamical} already establishes the pole structure that any such implementation must reproduce.

\section{Conclusion}\label{sec:conclusion}

Using exact diagonalization of the half-filled extended Hubbard model as a numerically exact reference, we have quantified the vertex correction demanded of the $GW$ approximation across the full correlation phase diagram. We found that the effective vertex changes character across the diagram: it is suppressive ($\Gamma_{\rm eff}<1$) in the weak-coupling regime, where RPA overestimates charge-fluctuation-driven gap renormalization, and enhancing ($\Gamma_{\rm eff}\gg 1$, growing $\sim U$) in the Mott regime, where it must compensate the missing derivative discontinuity. Vertex corrections in the electron-hole channel worsen the quasiparticle gap, indicating that reproducing the Mott gap requires dynamical self-energy structure absent from static GW. Static COHSEX is accidentally exact at a single crossover, $U^*\approx 3.5\,t$ for $V=0$, and we showed that non-local interactions split this point into a fragile double-crossover window at weak coupling. These results provide a quantitative, channel-selective diagnostic of $GW$ reliability and a clear directive for method development: in correlated materials, accuracy must be sought in the dynamical self-energy vertex, while the apparent success of static COHSEX at intermediate coupling should be treated as error cancellation rather than controlled accuracy.

\section*{Data Availability}
The exact-diagonalization and static-$GW$ code used to generate all data, figures, and tables in this work is openly available on Zenodo~\cite{AtamboZenodo2026} under the MIT license, together with a Jupyter notebook that reproduces every result reported here, including the $N=14$ finite-size extension of Appendix~\ref{app:finitesize}.

\begin{acknowledgments}
The author acknowledges Kenya Education Network (KENET) research services for computing resources.
\end{acknowledgments}


\begin{table}[htb]
\caption{Crossover points $U^*$ (in units of $t$) at which static $GW$ (COHSEX $\Gamma_{\rm eff}=1$) reproduces the exact gap, as a function of the nearest-neighbor interaction $V$. For $V=0$ a single accidental-cancellation point exists; for $V>0$ it splits into a double-crossover window $(U_1^*, U_2^*)$.}
\label{tab:crossover}
\begin{ruledtabular}
\begin{tabular}{ccc}
$V/t$ & $U_1^*/t$ & $U_2^*/t$ \\
\hline
0.0 & 3.5228 & - \\
0.5 & 0.7172 & 1.0420 \\
1.0 & 0.7685 & 1.4666 \\
\end{tabular}
\end{ruledtabular}
\end{table}

\clearpage
\appendix

\section{Finite-size scaling of the crossover}\label{app:finitesize}

To assess the robustness of the accidental cancellation, we extended the exact diagonalization to $N=10$ closed-shell clusters using sparse Lanczos techniques. Table~\ref{tab:gammasize} confirms that the qualitative behavior of $\Gamma_{\rm eff}$ (suppression at weak coupling, enhancement at strong coupling) is perfectly preserved across system sizes.

\begin{table}[htb]
\caption{Effective vertex $\Gamma_{\rm eff}$ versus $U/t$ for $N=6$, $10$, and $14$ ($V=0$). The crossover $\Gamma_{\rm eff}=1$ occurs at $U^* = 3.52\,t$ ($N=6$), $2.65\,t$ ($N=10$), and $2.27\,t$ ($N=14$). At fixed $U=8t$, $\Gamma_{\rm eff}$ grows approximately linearly with $N$ ($2.90,\ 5.33,\ 7.82$), i.e.\ without bound as the thermodynamic limit is approached.}
\label{tab:gammasize}
\begin{ruledtabular}
\begin{tabular}{cccc}
$U/t$ & $N=6$ & $N=10$ & $N=14$ \\
\hline
1 & 0.212 & 0.215 & 0.228 \\
2 & 0.426 & 0.574 & 0.750 \\
3 & 0.773 & 1.292 & 1.938 \\
4 & 1.224 & 2.272 & 3.523 \\
5 & 1.706 & 3.252 & 4.981 \\
6 & 2.159 & 4.091 & 6.154 \\
7 & 2.557 & 4.774 & 7.081 \\
8 & 2.896 & 5.328 & 7.822 \\
\end{tabular}
\end{ruledtabular}
\end{table}

\begin{figure}[htb]
\caption{Effective vertex $\Gamma_{\rm eff}(U)$ for $N=6,10,14$ ($V=0$). At fixed $U$ the required scalar renormalization grows systematically with system size, approximately linearly in $N$ at strong coupling, confirming that the static functional becomes progressively less adequate as the thermodynamic limit is approached.}
\label{fig:gamma-N}
\includegraphics[width=\columnwidth]{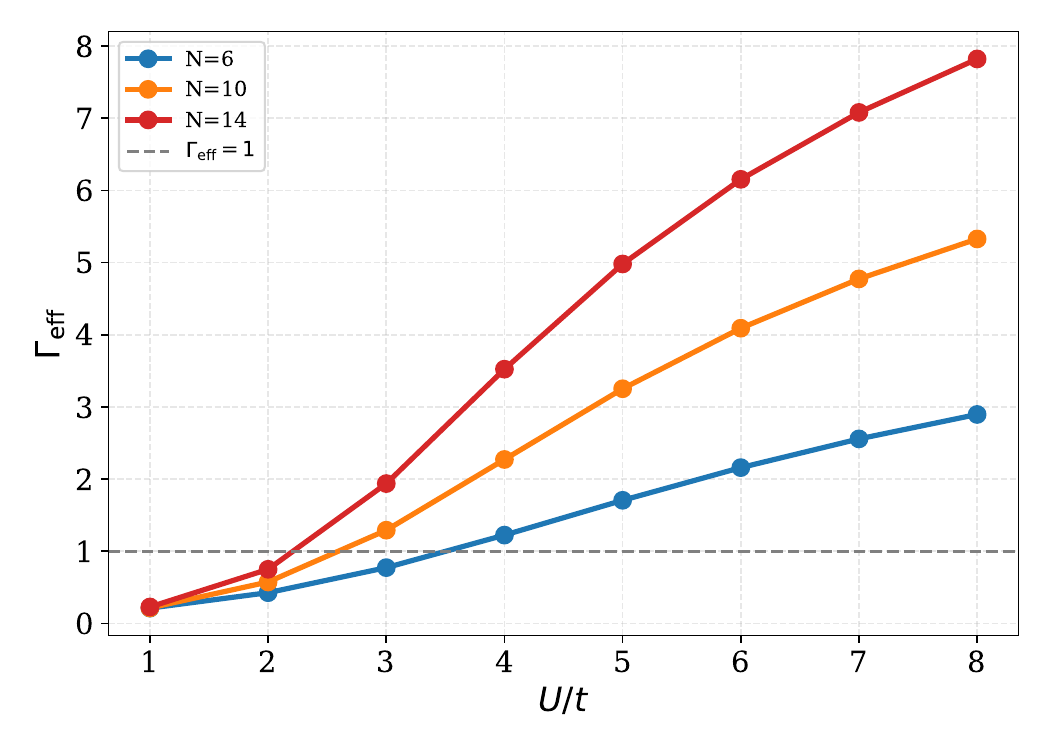}
\end{figure}

\begin{figure}[htb]
\caption{Finite-size drift of the accidental-cancellation point. Crossover $U^*(N)$ versus the finite-size non-interacting gap $\Delta_{\rm TB}(N)=4\sin(\pi/N)$ for $N=6,10,14$ (filled markers), with the analytic thermodynamic-limit endpoint $(0,0)$ shown as an open conceptual marker rather than a fitted extrapolation. Together with the unbounded growth of $\Gamma_{\rm eff}$ at fixed $U$ (Table~\ref{tab:gammasize}), the monotonic drift shows that the accidental-cancellation window shrinks toward $U=0$ as the finite-size level-spacing gap closes. No scaling law is claimed.}
\label{fig:scaling}
\includegraphics[width=\columnwidth]{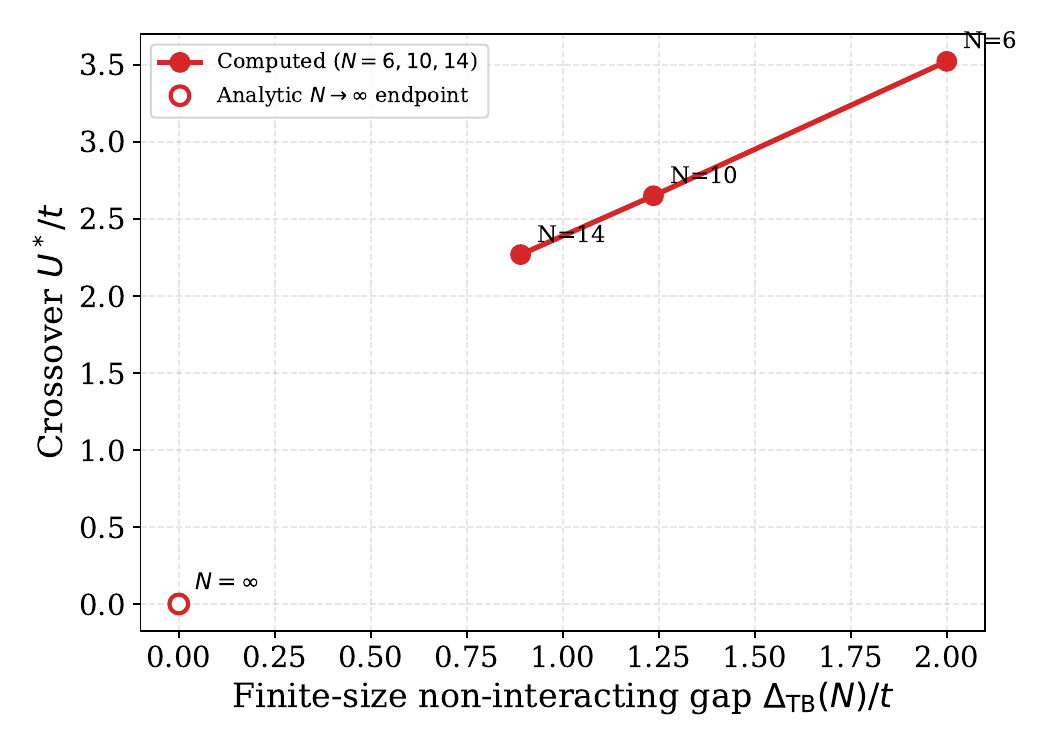}
\end{figure}

Crucially, the crossover point $U^*(N)$ drifts monotonically with cluster size: $U^* = 3.52\,t$ ($N=6$), $2.65\,t$ ($N=10$), and $2.27\,t$ ($N=14$), tracking the closure of the finite-size non-interacting gap $\Delta_{\rm TB}(N) = 4\sin(\pi/N)$ (Fig.~\ref{fig:scaling}). Equally important, at fixed strong coupling the required scalar renormalization grows without bound: $\Gamma_{\rm eff}(U=8t) = 2.90,\ 5.33,\ 7.82$ for $N=6,10,14$, approximately linear in $N$ ($\approx 0.62$ per site; Fig.~\ref{fig:gamma-N}). No finite static rescaling of the COHSEX self-energy can therefore reproduce the Mott gap in the bulk. We emphasize that the drift is monotonic but \emph{not} linear in $1/N$-a naive linear fit would misleadingly suggest a finite bulk crossover-so no extrapolation is performed. The thermodynamic-limit endpoint $U^* \to 0$ is instead fixed analytically: in the bulk the non-interacting half-filled chain is metallic and the static COHSEX self-energy constructed from a metallic reference remains gapless, whereas the exact model is insulating for any $U>0$~\cite{LiebWu1968}. As an internal check, a linear-in-$1/N$ extrapolation of the exact cluster gaps yields $\approx 1.0\,t$ at $U=4t$ and $\approx 4.6\,t$ at $U=8t$, the expected magnitude of the thermodynamic Lieb-Wu charge gap. The accidental cancellation is thus a finite-size artifact that vanishes in the thermodynamic limit.

\begin{thebibliography}{0}%
\makeatletter
\providecommand \@ifxundefined [1]{%
 \@ifx{#1\undefined}
}%
\providecommand \@ifnum [1]{%
 \ifnum #1\expandafter \@firstoftwo
 \else \expandafter \@secondoftwo
 \fi
}%
\providecommand \@ifx [1]{%
 \ifx #1\expandafter \@firstoftwo
 \else \expandafter \@secondoftwo
 \fi
}%
\providecommand \natexlab [1]{#1}%
\providecommand \enquote  [1]{``#1''}%
\providecommand \bibnamefont  [1]{#1}%
\providecommand \bibfnamefont [1]{#1}%
\providecommand \citenamefont [1]{#1}%
\providecommand \href@noop [0]{\@secondoftwo}%
\providecommand \href [0]{\begingroup \@sanitize@url \@href}%
\providecommand \@href[1]{\@@startlink{#1}\@@href}%
\providecommand \@@href[1]{\endgroup#1\@@endlink}%
\providecommand \@sanitize@url [0]{\catcode `\\12\catcode `\$12\catcode `\&12\catcode `\#12\catcode `\^12\catcode `\_12\catcode `\%12\relax}%
\providecommand \@@startlink[1]{}%
\providecommand \@@endlink[0]{}%
\providecommand \url  [0]{\begingroup\@sanitize@url \@url }%
\providecommand \@url [1]{\endgroup\@href {#1}{\urlprefix }}%
\providecommand \urlprefix  [0]{URL }%
\providecommand \Eprint [0]{\href }%
\providecommand \doibase [0]{https://doi.org/}%
\providecommand \selectlanguage [0]{\@gobble}%
\providecommand \bibinfo  [0]{\@secondoftwo}%
\providecommand \bibfield  [0]{\@secondoftwo}%
\providecommand \translation [1]{[#1]}%
\providecommand \BibitemOpen [0]{}%
\providecommand \bibitemStop [0]{}%
\providecommand \bibitemNoStop [0]{.\EOS\space}%
\providecommand \EOS [0]{\spacefactor3000\relax}%
\providecommand \BibitemShut  [1]{\csname bibitem#1\endcsname}%
\let\auto@bib@innerbib\@empty
\end{thebibliography}%


\begin{thebibliography}{99}
\bibitem{Strinati1980} G. Strinati, H. J. Mattausch, and W. Hanke, Phys. Rev. Lett. \textbf{45}, 290 (1980).

\bibitem{Strinati1982} G. Strinati, H. J. Mattausch, and W. Hanke, Phys. Rev. B \textbf{25}, 2867 (1982).

\bibitem{Godby1986} R. W. Godby, M. Schl\"uter, and L. J. Sham, Phys. Rev. Lett. \textbf{56}, 2415 (1986).

\bibitem{Hedin1965} L. Hedin, Phys. Rev. \textbf{139}, A796 (1965).

\bibitem{HedinLundqvist1970} L. Hedin and S. Lundqvist, Solid State Phys. \textbf{23}, 1 (1970).

\bibitem{HybertsenLouie1986} M. S. Hybertsen and S. G. Louie, Phys. Rev. B \textbf{34}, 5390 (1986).

\bibitem{AryasetiawanGunnarsson1998} F. Aryasetiawan and O. Gunnarsson, Rep. Prog. Phys. \textbf{61}, 237 (1998).

\bibitem{OnidaReiningRubio2002} G. Onida, L. Reining, and A. Rubio, Rev. Mod. Phys. \textbf{74}, 601 (2002).

\bibitem{BiermannAryasetiawanGeorges2003} S. Biermann, F. Aryasetiawan, and A. Georges, Phys. Rev. Lett. \textbf{90}, 086402 (2003).

\bibitem{KotaniSchilfgaarde2006} T. Kotani, M. van Schilfgaarde, and S. V. Faleev, Phys. Rev. Lett. \textbf{96}, 226402 (2006).

\bibitem{SchilfgaardeKotaniFaleev2006} M. van Schilfgaarde, T. Kotani, and S. Faleev, Phys. Rev. Lett. \textbf{96}, 226402 (2006).

\bibitem{Rohringer2018} G. Rohringer, H. Hafermann, A. Toschi, A. A. Katanin, A. E. Antipov, M. I. Katsnelson, A. I. Lichtenstein, A. N. Rubtsov, and K. Held, Rev. Mod. Phys. \textbf{90}, 025003 (2018).

\bibitem{AyralWernerBiermann2012} T. Ayral, P. Werner, and S. Biermann, Phys. Rev. Lett. \textbf{109}, 226401 (2012).

\bibitem{AryasetiawanCumulant2006} F. Aryasetiawan, M. Imada, A. Georges, G. Kotliar, S. Biermann, and A. I. Lichtenstein, Phys. Rev. B \textbf{74}, 195104 (2006).

\bibitem{DelSoleReiningGodby1994} R. Del Sole, L. Reining, and R. W. Godby, Phys. Rev. B \textbf{49}, 9898 (1994).

\bibitem{SchoneEguiluz1998} W.-D. Sch\"one and A. G. Eguiluz, Phys. Rev. Lett. \textbf{81}, 1662 (1998).

\bibitem{LiebWu1968} E. H. Lieb and F. Y. Wu, Phys. Rev. Lett. \textbf{20}, 1445 (1968).

\bibitem{GeorgesKotliar1996} A. Georges, G. Kotliar, W. Krauth, and M. J. Rozenberg, Rev. Mod. Phys. \textbf{68}, 13 (1996).

\bibitem{LimaCapelle2003} N. A. Lima, M. F. Silva, L. N. Oliveira, and K. Capelle, Phys. Rev. Lett. \textbf{90}, 146402 (2003).

\bibitem{AtamboZenodo2026} M. O. Atambo, \textit{Exact-diagonalization benchmarks for $GW$ and vertex corrections in the finite extended Hubbard ring (``Accidental Accuracy'')}, Zenodo (2026), \href{https://doi.org/10.5281/zenodo.22918065}{doi:10.5281/zenodo.22918065}.

\end{thebibliography}
\end{document}